\PassOptionsToPackage{sort&compress}{natbib}
\documentclass[article]{IEEEtran}
\usepackage{graphicx}  
\usepackage{dcolumn}   
\usepackage{bm}        
\usepackage{amssymb}   
\usepackage{amsmath}
\usepackage{bbm,amsfonts}

\usepackage{color}
\usepackage{amsthm}
\usepackage{setspace}
\usepackage{nomencl}
\usepackage{amstext} 
\usepackage{siunitx}
\usepackage{subfigure}

\usepackage{tabularx}
\usepackage[utf8]{inputenc}
\usepackage[dvipsnames]{xcolor}

\usepackage{comment}
\usepackage{physics}

\usepackage{url}
\usepackage{breakurl}
\usepackage[breaklinks]{hyperref}
\hypersetup{colorlinks}

\title{Demonstration of quantum volume 64 on a superconducting quantum computing system}

\author{\IEEEauthorblockN{Petar Jurcevic\IEEEauthorrefmark{1}, Ali Javadi-Abhari\IEEEauthorrefmark{1}, Lev S. Bishop\IEEEauthorrefmark{1}, Isaac Lauer\IEEEauthorrefmark{1}, Daniela F.~Bogorin\IEEEauthorrefmark{1}, Markus Brink\IEEEauthorrefmark{1}, Lauren Capelluto\IEEEauthorrefmark{1}, Oktay G\"{u}nl\"{u}k\IEEEauthorrefmark{1}, Toshinari Itoko\IEEEauthorrefmark{2}, Naoki Kanazawa\IEEEauthorrefmark{2}, Abhinav Kandala\IEEEauthorrefmark{1}, George A.~Keefe\IEEEauthorrefmark{1}, Kevin Krsulich\IEEEauthorrefmark{1}, William Landers\IEEEauthorrefmark{1}, Eric P.~Lewandowski\IEEEauthorrefmark{1}, Douglas T.~McClure\IEEEauthorrefmark{1}, Giacomo Nannicini\IEEEauthorrefmark{1}, Adinath Narasgond\IEEEauthorrefmark{1}, Hasan M. Nayfeh\IEEEauthorrefmark{1}, Emily Pritchett\IEEEauthorrefmark{1}, Mary Beth Rothwell\IEEEauthorrefmark{1}, Srikanth Srinivasan\IEEEauthorrefmark{1}, Neereja Sundaresan\IEEEauthorrefmark{1}, Cindy Wang\IEEEauthorrefmark{1}, Ken X.~Wei\IEEEauthorrefmark{1}, Christopher J. Wood\IEEEauthorrefmark{1}, Jeng-Bang Yau\IEEEauthorrefmark{1}, Eric J.~Zhang\IEEEauthorrefmark{1}, Oliver E.~Dial\IEEEauthorrefmark{1}, Jerry M.~Chow\IEEEauthorrefmark{1}, Jay M.~Gambetta\IEEEauthorrefmark{1}}

\IEEEauthorblockA{\IEEEauthorrefmark{1}IBM Quantum,
IBM T.J. Watson Research Center, Yorktown Heights, NY 10598, USA}

\IEEEauthorblockA{\IEEEauthorrefmark{2}IBM Quantum, IBM Research Tokyo, 19-21 Nihonbashi Hakozaki-cho, Chuo-ku, Tokyo, 103-8510, Japan}
}

\date{August 2020}

\begin{document}

\maketitle
\begin{abstract}
We improve the quality of quantum circuits on superconducting quantum computing systems, as measured by the quantum volume, with a combination of dynamical decoupling, compiler optimizations, shorter two-qubit gates, and excited state promoted readout. This result shows that the path to larger quantum volume systems requires the simultaneous increase of coherence, control gate fidelities, measurement fidelities, and smarter software which takes into account hardware details, thereby demonstrating the need to continue to co-design the software and hardware stack for the foreseeable future.   
\end{abstract}
\section{Introduction}

Quantum computing is a new kind of computing, using the same physical rules that atoms follow in order to manipulate information. At this fundamental level, quantum computers execute quantum circuits -- like a classical computer's logical circuits -- but now using the physical phenomena of superposition, entanglement, and interference to implement mathematical calculations that are out of reach for even our most advanced supercomputers.\\

As we progress towards machines capable of implementing circuits with a quantum advantage, meaning certain information processing tasks can be performed more efficiently or cost effectively than with classical circuits, quantum volume (QV)~\cite{Cross2019} serves as a holistic benchmark for quantum systems indicating the size of the quantum circuits that can be run on them. Sensitive to improvements in many aspects of device performance, quantum volume includes gate errors, measurement errors, the quality of the circuit compiler, and spectator errors. In Ref.~\cite{Cross2019} and later in Ref.~\cite{Pino2020}, QV16 was measured on \emph{ibmq\_johannesburg} and a Honeywell quantum system, respectively. In Ref.~\cite{Karalekas_2020}  QV8 was measured for the Rigetti Aspen-4 quantum system. We recently increased \emph{ibmq\_johannesburg} to QV32~\cite{Sundaresan2020} by improving our physical understanding of the two-qubit cross-resonance gate and using rotary echo pulses to reduce gate and spectator errors. Finally in unpublished work Honeywell has claimed to measure QV64~\cite{Honeywell64}.\\ 

Here we demonstrate an increase in the quantum volume of an IBM quantum system by improving the Qiskit compiler~\cite{Qiskit}, implementing excited state promoted (ESP) readout, shorter two-qubit gates, and adding dynamic decoupling to the idle qubits. These last three demonstrate the need for timing and pulse control in cloud quantum systems~\cite{mckay2018qiskit}. While individually not one of these improvements is enough to allow \textit{ibmq\_montreal} to reach QV64, when combined we achieve QV64 with a heavy output probability (HOP) of $0.701 \pm  0.031(> 2/3 \pm 2\sigma)$ with a confidence interval of $98.744\%$ $(z=2.25)$, see Fig.~\ref{fig:HOPSuccess}(a). The quantum volume test requires exceeding 2/3 HOP by a $97.725\%$ $(z=2)$ confidence interval.\\

In Section~\ref{sec:dev} we give an overview of the \emph{ibmq\_montreal} device, which is a 27-qubit IBM Quantum Falcon processor; in Section~\ref{sec:com} we discuss the improvements to the compiler; in Section~\ref{sec:DD} we discuss the dynamical decoupling protocol; in Section ~\ref{sec:faster-cx} we discuss the faster implementation of the direct CNOT gate which extends the improved pulse control of~\cite{Sundaresan2020}; in Section~\ref{sec:esp} we discuss the improvement in measurement fidelity by using a control pulse to promote the excited state to a higher level before measurement~\cite{Mallet2009}. Finally in Section~\ref{sec:con} we conclude the paper.

\section{Quantum System - \emph{ibmq\_montreal}}\label{sec:dev}

\begin{figure}[h]
\includegraphics[width=8.6cm]{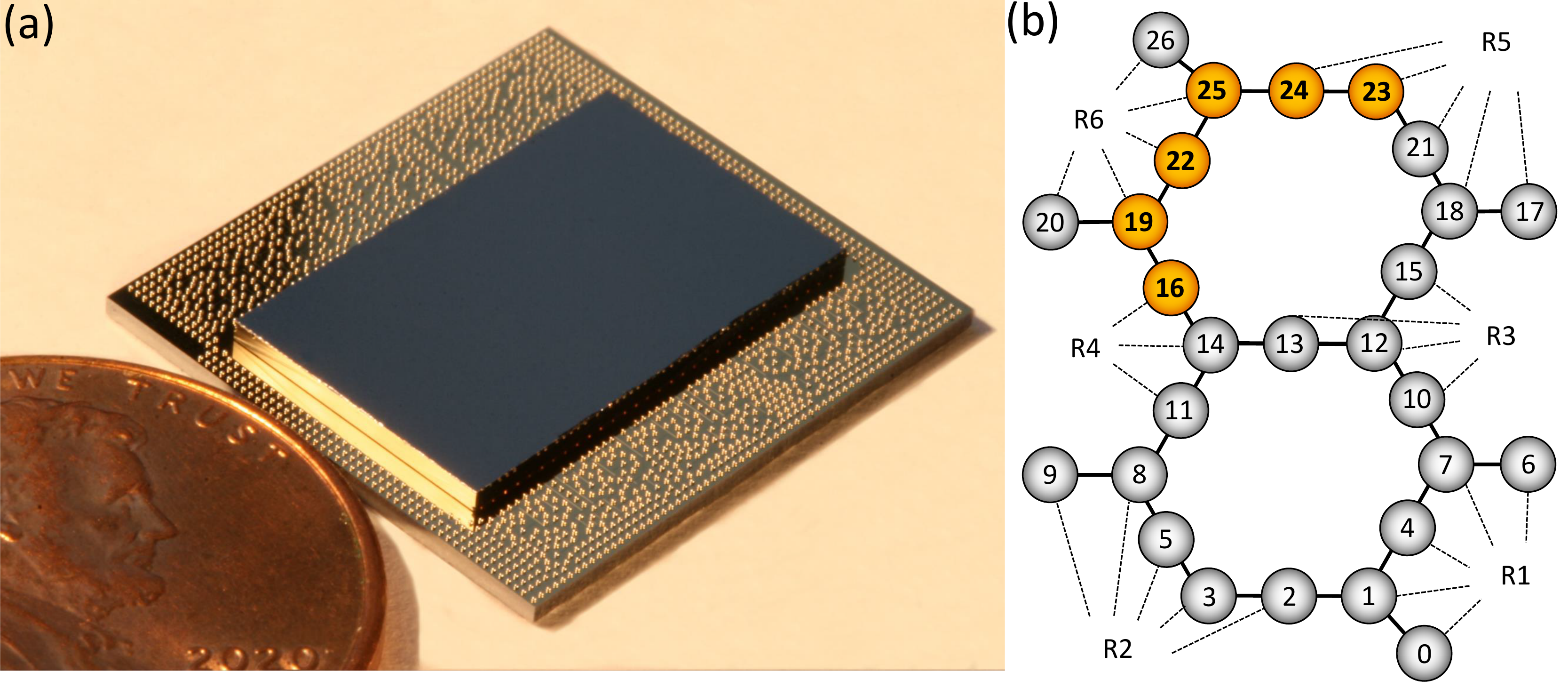}
\caption{\label{fig:devices} (a) Image of a representative IBM Quantum Falcon processor with a penny for scale. The lattice connectivity is defined through couplings on a top qubit die which is bump-bonded to a bottom interposer die for signal delivery and readout. (b) Schematic of the 27-qubit (numbered 0 through 26) heavy-hex layout connectivity. Qubits used for the confirmed QV64 are shaded in orange. Dashed lines indicate collections of qubits that are multiplexed together for readout (labeled R1 to R6).}
\end{figure}

The device studied in this work is from the recent series of IBM Quantum Falcon processors, which consist of 27 qubits arranged in a lattice designed for a distance-3 hybrid Bacon-Shor-surface code~\cite{Chamberland2019}. A photo of this processor is shown in Fig. \ref{fig:devices}(a), and a schematic of its connectivity is shown in  Fig.~\ref{fig:devices}(b). A high connectivity layout, such as `all-to-all', is preferable for random quantum circuits (such as QV circuits) in order to minimize the average qubit-qubit distance; however, additional edges in the connectivity increase the chance of frequency collision, cross-talk, and spectator errors. The IBM Quantum Falcon processor is a compromise, preserving a connectivity efficient for a logical qubit while simultaneously reducing detrimental effects of collisions and cross-talk without excessive insertion of SWAP gates to emulate `all-to-all' connectivity. In these and related systems, by using the techniques described in~\cite{Sundaresan2020}, we have measured a QV of 32 on the last 7 deployed systems~\cite{IQX} demonstrating the reliability of this architecture. \\

\begin{figure}

\includegraphics[width=8.6cm]{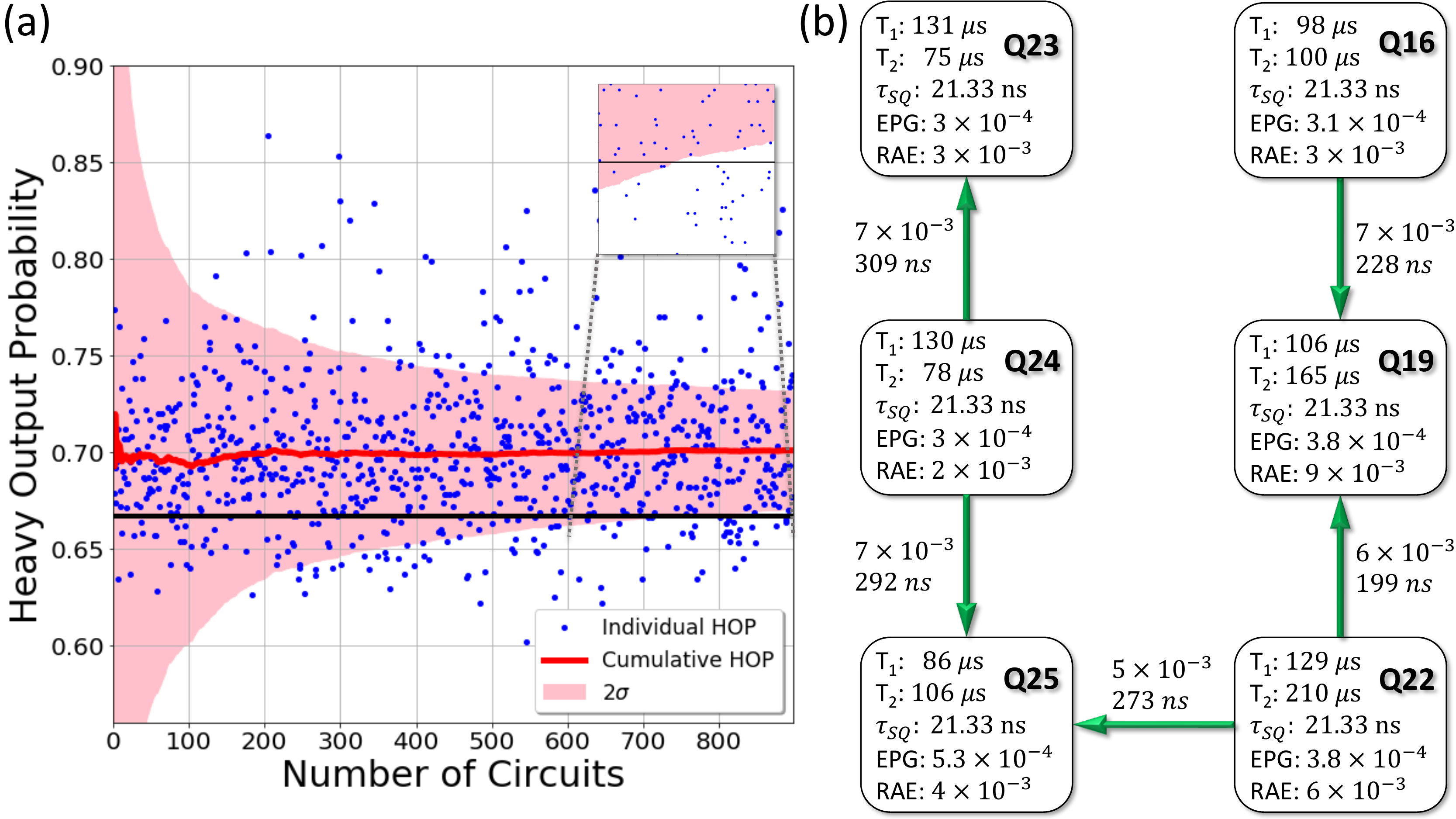}
\caption{\label{fig:HOPSuccess} (a) One of two statistically confirmed QV64 runs.  Here, a total of $\approx 900$ random circuits are run. Inset: QV success criteria were reached  $ > 724$ circuits. Blue: Heavy output probability for each individual circuit. Red: Cumulative heavy output probability with shaded region $\pm 2\sigma$ as calculated per appendix C of~\cite{Cross2019}. Black: Quantum volume success threshold at $2/3$. (b) Qubits used in the successful QV64 measurement. EPG: error per gate (single-qubit) measured with RB. RAE: readout assignment error. $\tau_{SQ}$: single-qubit gate duration. Natural two-qubit gate direction is shown in green, from control to target. Two-qubit gate error rates and gate durations are shown next to the corresponding qubit-qubit link.}
\end{figure}

In this paper we achieve QV64 on \textit{ibmq\_montreal}, which is one of the latest deployed IBM Quantum Falcon processors. The quantum volume circuits were run on a line of six qubits, Q16-Q19-Q22-Q25-Q24-Q23 (orange shaded qubits in Fig.~\ref{fig:devices}(b). Individual qubit properties are shown in Fig~\ref{fig:HOPSuccess}(b) with the following average values: $\mathrm{T}_1= \SI{113}{\mu\second}$, $\mathrm{T}_2= \SI{122}{\mu\second}$, error per single-qubit gate $3.8\times10^{-4}$, error per two-qubit gate $6.4\times10^{-3}$, and single-qubit readout assignment error $6.0\times10^{-3}$. Gate errors were measured with simultaneous single-qubit and individual two-qubit randomized benchmarking~\cite{Gambetta2012}.\\

The qubits are fixed-frequency transmons with frequencies $\approx \SI{5}{\giga\hertz}$. Single-qubit gates are driven resonantly with a microwave pulse of duration $\mathrm{\tau_{sq}} = \SI{21.33}{n\second}$. A DRAG pulse envelope~\cite{Motzoi2009} corrects $\sigma_z$-errors and signal dispersion due to wiring. Two-qubit gates are based on a cross-resonance scheme~\cite{Paraoanu2006,Rigetti2010,Chow2011} with a target rotary pulse~\cite{Sundaresan2020} and an additional offset pulse-shape on the target for implementing a direct (echoless) CNOT as described later. Two-qubit gate lengths are $\mathrm{\tau_{tq}} = \SI{199}-\SI{309}{\nano\second}$.
 
\section{Compiler}\label{sec:com}

Circuit compilation is a substantial part of quantum computation. Here we report improvements in the state-of-the-art Qiskit compiler to achieve reductions in the number of gates which results in circuits with shorter depths.
The compilation of a quantum volume circuit for a superconducting processor can be roughly broken down into two stages. The first stage is to map the circuit to the hardware's qubit connectivity constraints. At the conclusion of this step, each circuit will consist of a series of SU(4) gates on the available links, as well as the overhead of routing qubit information on the physical fabric, usually in the form of SWAPs. The second step consists of local expansions to the native gates of the hardware and optimizations. We introduce new compiler passes to improve both stages, and leverage existing passes in the Qiskit compiler throughout to achieve further reductions where possible: approximate synthesis, commutative cancellation, and peephole optimization of single-qubit and two-qubit chains of gates.\\

It is worth noting that the particular passes reported here have general utility beyond QV\@. Qubit mapping and routing is ubiquitous in compiling for  limited-connectivity architectures, and SU(4) synthesis has broad use in peephole optimization of sequential two-qubit gates.\\

\paragraph{Qubit layout and routing via Binary Integer Programming}
We formulate qubit layout and routing as a binary integer programming (BIP) problem, which we are able to solve to optimality. We choose as the cost function, $C$, the effective fidelity, modeled as the product of the fidelity of all the implemented gates:
\begin{align}
    C=K^d\prod_{j\in G} F_j^\text{best}\prod_{j\in \bar{G}} \bar{F}_j^\text{best}\prod_{j\in S} F_b^3 ,
\end{align}
where $K$ is a factor penalizing circuits with high depth $d$;  $G$ [$\bar{G}$] are the set of gates that are mapped directly [mapped with mirroring -- combining SWAP with a gate]; and $S$ is the set of added SWAP gates. Here, $F_b$ is the gate fidelity of the available entangling gate (which must be applied 3 times to implement SWAP), $F_j^\text{best}$ [$\bar{F}_j^\text{best}$] is the modeled fidelity of the best approximation to the target unitary making $i=0,\ldots, 3$ uses of the entangling gate
\begin{align}
    F_j^\text{best} &= \max_i F^\text{avg}_{i,j} (F_b)^i ,\\
    \bar{F}_j^\text{best}&=\max_i \bar{F}^\text{avg}_{i,j} (F_b)^i ,
\end{align}
and $F^\text{avg}_{i,j}$ is the average gate fidelity due to approximating the $j$-th gate with $i$ uses of the entangling gate \cite[Appendix B]{Cross2019}.\\

\begin{figure}
\includegraphics[width=8.6cm]{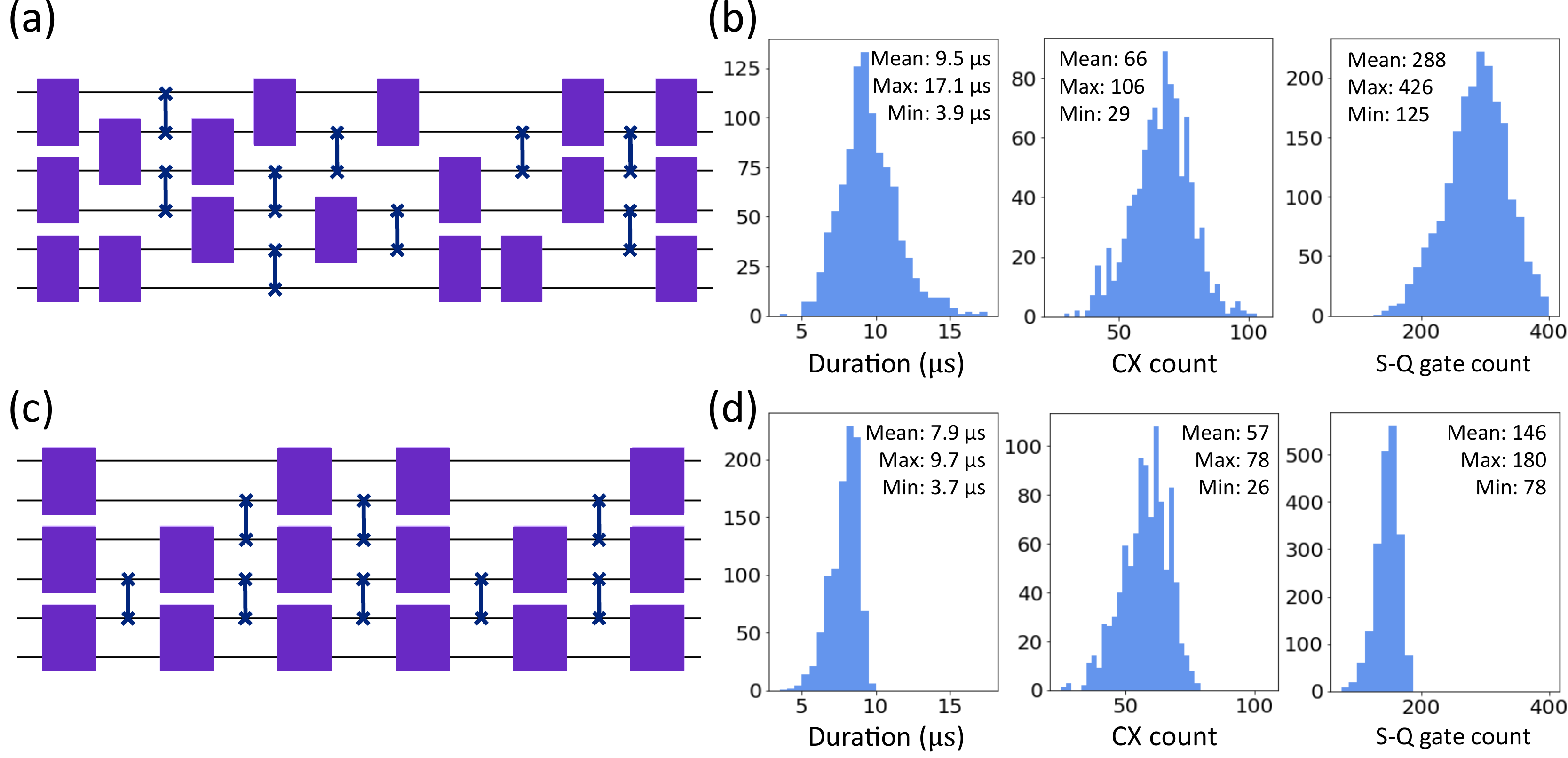}
\caption{\label{fig:compiler}Comparison of QV64 circuits transpiled to a line connectivity with (a), (b) the state-of-the-art Qiskit compiler and (c), (d) an improved transpilation method based on BIP and additional gate cancellations, see text. (a) and (c) show the same random example circuit, mapped with ``SABRE" and mapped with BIP, respectively. The purple boxes represent the random SU(4) and SWAPs are indicated in blue. (b) and (d) show statistics of 2000 circuits using both methods. CX count improvements are due to improved mapping, and S-Q (single-qubit) count improvements the result of pulse-efficient compilation. Both contribute to shorter durations. We assume basis gate fidelity $F_b=0.99$ for the approximate SU(4) expansion in all cases. If the native gate is ECR (rather than direct-CX), we get additional 7\% reduction in mean duration by targeting the native gate and absorbing local pre-rotations.}
\end{figure}

The freedom to implement either a gate or its mirror allows elimination of many explicit SWAP gates, and by restricting the number of candidate SWAP insertion sites we are able to reduce the size of the BIP problem such that it can be solved to optimality in around one second per circuit, using optimization software such as CPLEX~\cite{cplex2009v12}.
Figure~\ref{fig:compiler} shows the performance of this BIP pass in comparison to the state-of-the art SABRE algorithm~\cite{li_tackling_2019} available in Qiskit, showing substantial improvement in both the mean and maximum number of uses of the entangling gate.\\ 

\paragraph{Pulse-efficient SU(4) decomposition} The \emph{ibmq\_montreal} device has the following native gate set for achieving universal quantum computation: Ctrl-X (CX), Sqrt-X (SX) and Phase($\theta$).
The CX gate itself can be implemented directly or be created using an Echo Cross-Resonance (ECR) pulse\cite{Sheldon2016} (c.f. Section~\ref{sec:faster-cx}). The Phase gate can be achieved with zero time and error~\cite{McKay2016}. We refer to any gate that is one pulse (i.e. equivalent to an SX by a pre-/post-phase) as a single-qubit (SQ) gate (e.g. Hadamard). A generic single-qubit operation (U) can be achieved with at most 2 SQ pulses.\\

\begin{figure}
\includegraphics[width=8.9cm]{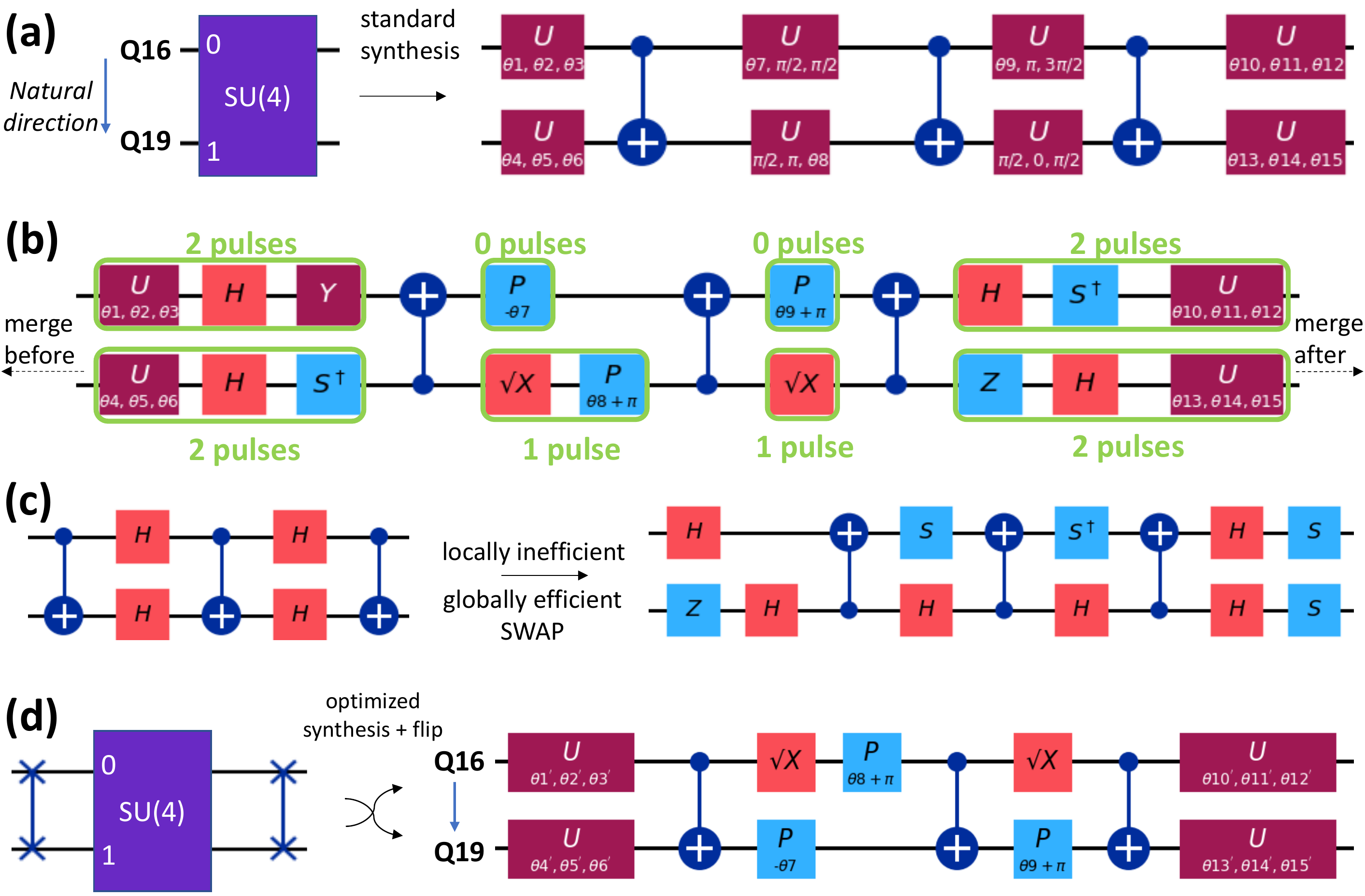}
\caption{\label{Singles}
(a) Standard (Qiskit) decomposition of an SU(4) operator in terms of 3 CNOTs and layers of single-qubit rotations. Each U contributes 2 SQ pulses for a total of 16 pulses. When the expansion is not in the ``natural" direction of the hardware CX, extra single qubit rotations will be  involved.
(b) A new pulse-efficient SU(4) decomposition obtained constructively from the first (c.f. same 15 $\theta$ parameters).
(c) This decomposition applied to SWAPs creates more global efficiency. Even though more pulses are used locally (H), the ``inner" pulses are reduced and the ``outer" ones can merge with gates before and after the SWAP.
(d) If the direction of the above expansion is incorrect, synthesize a different ``doubly-mirrored" operator, then flip it at the point of use. This has the same number of pulses but now in the correct direction.}
\end{figure}

Given the {CX, SQ} or {ECR, SQ} set of native pulses, we aim to minimize them during the expansion of each SU(4) and SWAP. A second goal is to expand them in a way that creates further opportunities for optimization.
It is known that any SU(4) can be implemented using at most 3 CX gates~\cite{vatan2004optimal}, and 2 CX gates suffice for many useful approximations (e.g. at 99\% fidelity)~\cite{Cross2019} (cf. Figure~\ref{Singles}(a)). ECR is locally equivalent to CX, so it has the same requirements. While the question of ``optimal'' SU(4) decomposition has been extensively studied, the optimality criteria has usually been the number of 2-qubit gates~\cite{vatan2004optimal,shende2004minimal}. To extract ultimate performance, we are also interested in minimizing the total number of pulses and the duration.\\

Our approach is based on three strategies:

1. Circuit simplification to reduce redundant pulses: starting from a Qiskit synthesis of an arbitrary SU(4), we apply repeated circuit identities to the result to reduce its cost. This gives us a constructive SU(4) decomposition, depicted in Figure~\ref{Singles}(b), which is optimal in the number of pulses (by a simple parameter counting argument). This decomposition has another advantage, in that 8 out of 10 single-qubit pulses are placed on the outside of the structure. Given that 2 SQ pulses suffice for any aggregate single-qubit operation, this creates an opportunity for merging with preceding and following layers of SU(4) in the circuit.
One surprising consequence of this decomposition is that for the special case of a SWAP operation, the decomposition is locally less efficient than a textbook expansion; however globally it is more efficient as it creates more opportunities for cancellation (Figure~\ref{Singles}(c)).
We arrive at similar pulse-efficient decompositions targeting the ECR gate, and also for approximated SU(4)s that use 2 CX instead of 3 (omitted for brevity).\\

2. Decomposition in the natural gate direction: While the device software is easily capable of implementing a CX gate in both directions, in reality there is a preferred gate direction in terms of speed and error on the hardware. The other direction is achieved by local pre- and post-rotations. The same is true for ECR gates. By querying the device for its natural direction, we can expand each SU(4) and SWAP in the correct direction in the compiler, avoiding further cost down the road. To synthesize a general SU(4) when the logical and physical directions are mismatched, we employ a trick of double mirroring (adding SWAPs before and after the SU(4)). The doubly-mirrored SU(4) implements a different operator, where the middle two rows and middle two columns are swapped. We perform a pulse-efficient synthesis on the doubly-mirrored operator, but apply it in the circuit with the reverse order of qubits. This will ensure the original operator is implemented, but also now with the correct physical gate direction (Figure~\ref{Singles}). Double-mirroring creates a locally equivalent gate, so any approximation to the original SU(4) still holds with the same error bounds.\\

3. Decomposition to native gate: If a direct CX is not available, we compile to the fundamental two-qubit interaction available. In the case of ECR, this saves us the extra single-qubit pulses involved in creating a CX. This demonstrates the benefit of removing simplifying abstraction barriers in the exposed gate set to gain efficiency in compiling~\cite{maslov2017basic,murali2019full,arute2020quantum}.

\section{Dynamical Decoupling}
\label{sec:DD}
When quantum circuits are mapped to physical hardware, not all physical gates can be performed simultaneously.
Gate execution-times can vary significantly, not only between single- and two-qubit gates, but also between individual qubits and qubit-pairs. In addition, architecture-specific gate schemes and connectivity determine which and how many gates can be executed in parallel.\\  

\begin{figure*}
\includegraphics[width=\textwidth]{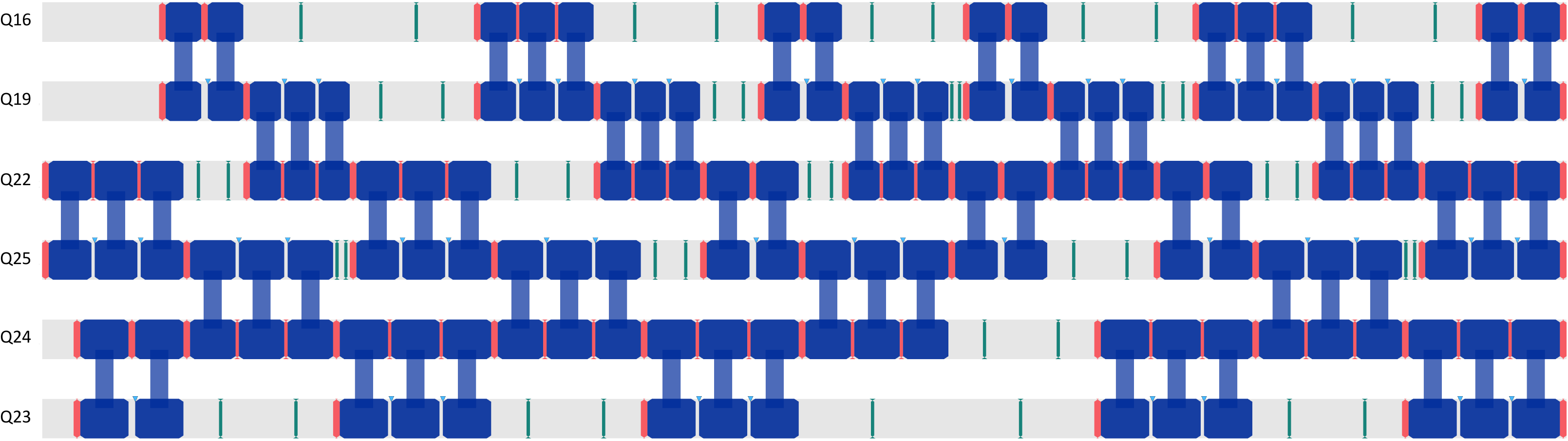}
\caption{\label{fig:sched_circ}An example QV64 circuit drawn as scheduled on the device. Two-qubit gates are depicted in blue, single-qubit gates in red with scaling proportional to their gate lengths. Grey areas indicate idle times on particular qubits. Dynamical decoupling pulses, in green, are placed symmetrically within idle times. Idle times range from half to six times a two qubit gate length.}
\end{figure*}

An analysis of QV64 circuits mapped to a line of transmon-qubits reveals idle times that are a significant portion of the total circuit duration (Fig.~\ref{fig:sched_circ}). Two main effects create these idle slots. Firstly, a line configuration with nearest-neighbor gates requires a total of 7.3 SWAPs on average per QV circuit. In the optimal layout and routing choice (Section~\ref{sec:com}) for reducing the number of SWAP gates, SWAPs are not executed at once over the entire quantum register, as shown in Fig.~\ref{fig:compiler}(c). When the basis two-qubit gate is a local equivalent of CNOT, this creates ``idle holes'' for the duration of three two-qubit gates  (Fig.~\ref{fig:sched_circ}).
Secondly, ``idle holes'' can still arise even if no SWAP operations are required. While single-qubit gates are tuned with identical durations across the entire register, two-qubit gate durations depend on qubit frequencies and coupling, differing by a factor of $1.5-2$ between the fastest and the slowest gates.  Given that two-qubit gates are $\approx 10 \times$ longer than single-qubit gates, these differences accumulate over the course of the computation, opening up additional temporal gaps when individual qubits sit idle. \\

Ideally idle qubits would evolve the identity operation; however, this is executed far from perfectly in realistic architectures. While thermal relaxation and white noise dephasing lead to dissipative information loss, cross-talk and unwanted non-local spectator interactions lead to local and non-local unitary errors, respectively. In addition, non-Markovian noise sources such as charge noise lead to non-white dephasing.  All three error sources are detrimental as circuits become larger, i.e., wider and deeper. Dynamical decoupling is a thoroughly discussed error mitigation technique ~\cite{Lorenza1999,Souza2012,Suter2016}, and in its simplest form, can be a single Hahn echo-pulse~\cite{Hahn1950}, refocusing the low-frequency noise spectrum acting on a unitary. Various decoupling sequences have been proposed~\cite{Carr1954,Meiboom1958,Maudsley1986}, some with self-correcting properties~\cite{Khodjasteh2005,Khodjasteh2007}, others with non-equidistant temporal spacing~\cite{Uhrig2007}, and hybrids combining both~\cite{Uhrig2009,Souza2011,Quiroz2011}, in order to optimize the effective filter function. Recently, dynamical decoupling has been shown to improve single-qubit states and an entangled two-qubit state on a Rigetti and IBM quantum computer \cite{Pokharel2018}.\\

For the successful QV64 measurement presented here, we used the sequence $\tau^{i,q}/2\, -\, X_p\, -\, \tau^{i,q}\, -\,X_m\, -\, \tau^{i,q}/2$, with delays $\tau^{i,q}=\left(T^{i,q}_{\mathrm{idle}}-2*T_{X_{p/m}}\right)/2$, where $T^{i,q}_{\mathrm{idle}}$ is the $i\mathrm{th}$ idle length on qubit $q$, and $T_{X_{p/m}}$ is the duration of one echo pulse with $X_{p,m}$ being a $\pi$-pulse around x-axis with positive/negative sense of rotation.

\begin{figure}
\makebox[8.6cm][c]{\includegraphics[width=5.5cm]{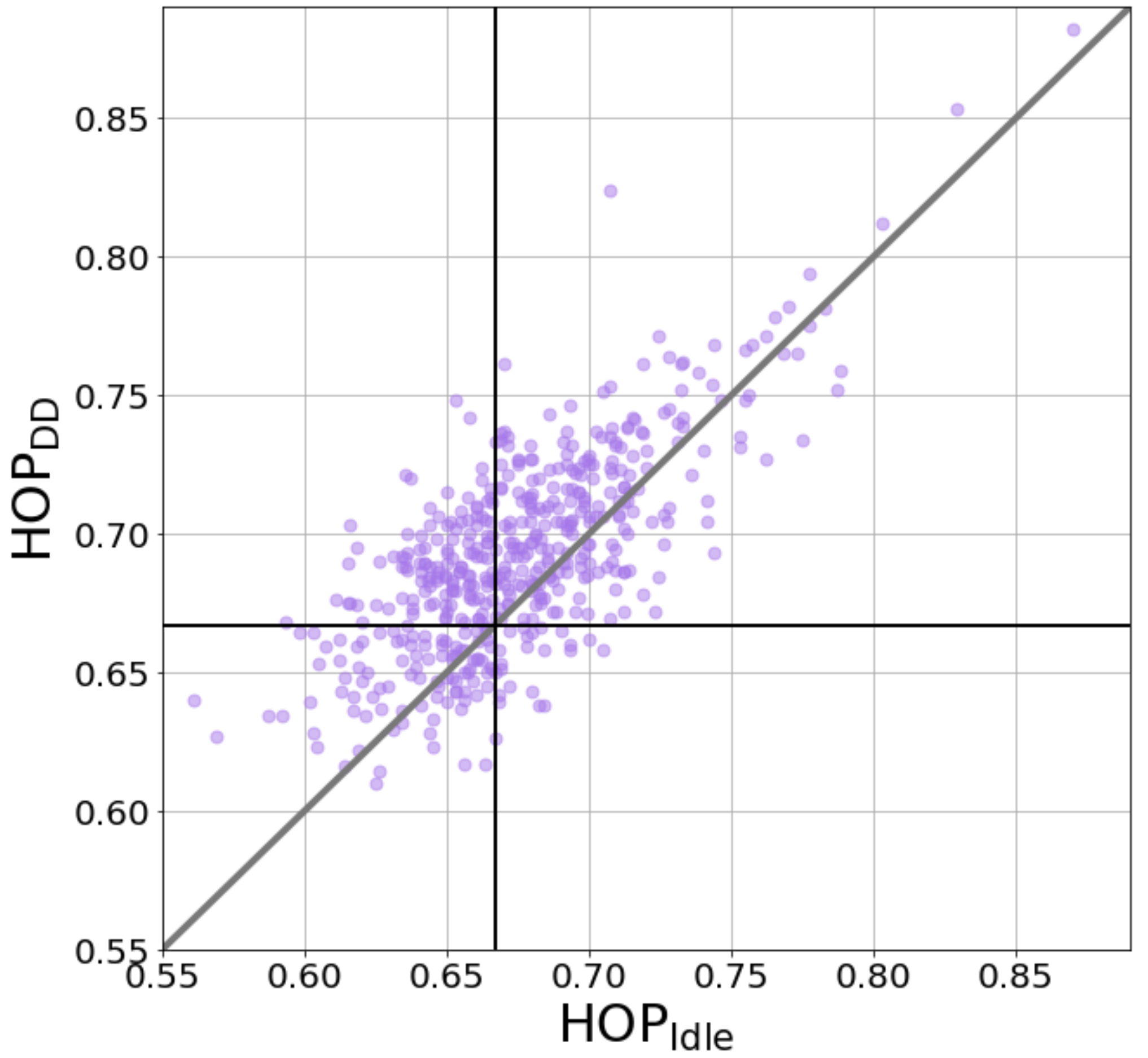}}
\caption{\label{fig:DD}Comparison between heavy output probabilities for the same circuits with and without dynamical decoupling. $2/3$-threshold is indicated by the horizontal (vertical) black line. $72.8\%$ of all circuits show larger HOP with decoupling, i.e. are above the grey line, with an average increase of $0.0178$.}
\end{figure}

Figure~\ref{fig:DD} shows a comparison of identical QV-circuits run with (DD) and without (Idle) dynamical decoupling. Dynamical decoupling  with $X_p-X_m$ sequences improves $72.8\%$ of all circuits in this run, i.e. $\mathrm{HOP_{DD}}>\mathrm{HOP_{Idle}}$, with an average HOP increase of $0.0178$. The interplay between various DD sequences and random circuits, such as QV circuits, is an open research focus. 

\section{Direct CX gate}
\label{sec:faster-cx}
Even with state-of-the-art compiling, QV64 circuits consist of a total of 57 two-qubit gates and 146 single-qubit gates on average. Any improvement in gate speed can significantly reduce the circuit duration compared to the coherence times. However, the optimal gate speed for running a circuit is in general not the speed that maximizes the fidelity of the individual gates. In particular, qubits experience idle times in a multi-qubit circuit (see Section~\ref{sec:DD}), and the fidelity of the identity operation during these idle times is not captured in the single-qubit or two-qubit randomized benchmarking fidelities often used to characterize quantum systems. Finding the optimal trade-off between individual gate fidelity and circuit fidelity is currently open research, in addition to characterizing which errors are enhanced by  driving gates faster.  Here we focus on techniques to reduce two-qubit gate durations, but note that small increases in the speed of either single- or two-qubit gates can significantly impact the performance of QV64 circuits. \\

\begin{figure}
\includegraphics[width=8.6cm]{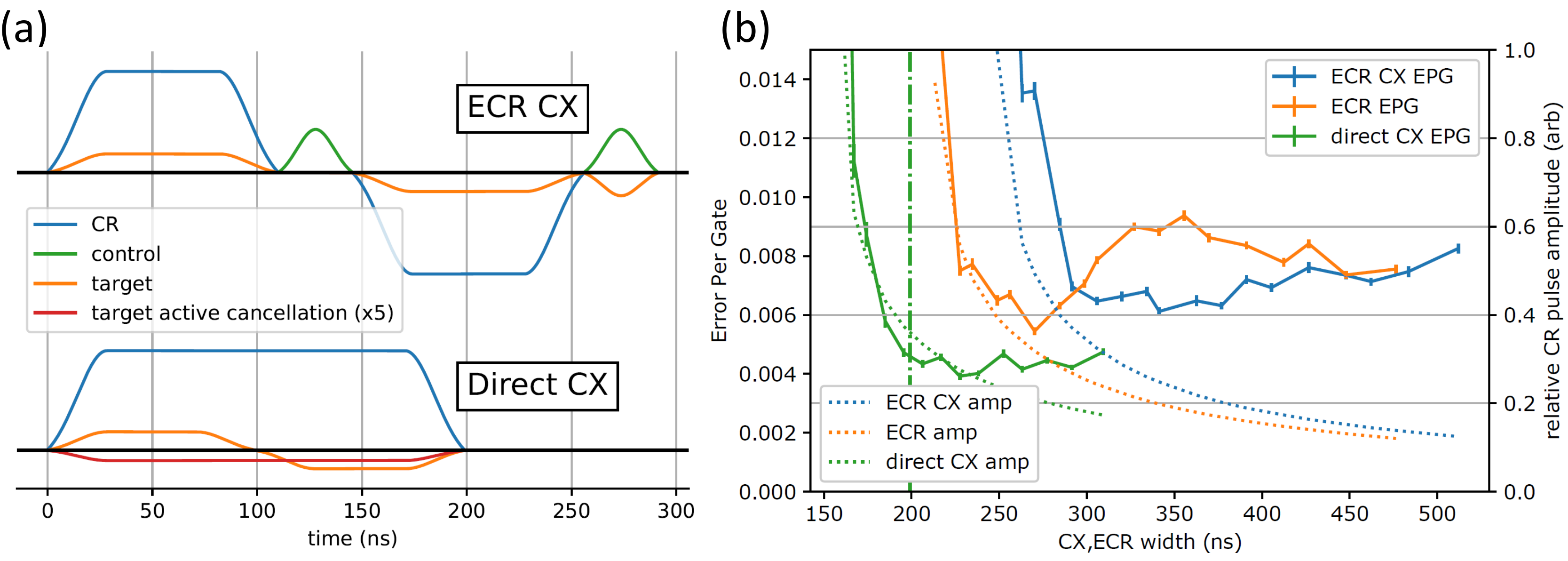}
\caption{\label{ddcx} (a) Pulse envelope comparison between the echoed cross resonance (ECR) CX gate
and the direct CX gate implementation of Control-Target C$_{22}$-T$_{19}$ on \textit{ibmq\_montreal}. (b) Error per gate vs. gate width for ECR CX (blue), ECR (orange), and direct CNOT (green). CR-drive signal amplitudes for the various gate versions and gate widths are shown by the dotted lines. Vertical dashed line indicates the direct CX gate width used for QV64.}
\end{figure}

As mentioned in Section~\ref{sec:com}, an immediate way to ``speed up" two-qubit gates is to incorporate into the circuit compilation any pre-/post-single-qubit rotations needed to get from the native ECR gate to a CX or CNOT.  We compare the standard echoed cross-resonance gate ECR CX, shown at the top of  Fig.~\ref{ddcx}(a), to an ECR gate in which single qubit rotations are compiled separately, reducing the two-qubit gate duration to only the entangling portion of the gate.  The errors of ECR CX and ECR, measured by two-qubit randomized benchmarking, are shown in Fig.~\ref{ddcx}(b) as a function of the two-qubit gate duration.   \\

Two-qubit gates can be further sped up by finding high-fidelity alternatives to the echo pulse sequence, effectively removing another single-qubit gate from the total two-qubit gate duration. We compare an example of a ``direct" echo-free CX pulse sequence, shown at the bottom of  Fig.~\ref{ddcx}(a), to ECR and ECR CX. 
This sequence demonstrates an improvement over previous direct CNOT attempts~\cite{Chow2011} by leveraging our understanding of target rotary pulsing~\cite{Sundaresan2020}. 
The resonant drive of the target is implemented as the sum of two parts, an active cancellation tone and a target rotary tone that are symmetric and antisymmetric over the CR pulse, respectively.  The active cancellation tone cancels IX terms in the native CR Hamiltonian and any IY terms due to classical crosstalk, while the target rotary pulse can be used to reduce unwanted ZZ and ZY.\\

The impact of reducing the total gate duration is clearly evidenced by a reduction of two-qubit gate error, as shown in Fig.\ref{ddcx}(b). All gate sequences -- ECR CX, ECR,  and direct CX -- experience a sudden loss of fidelity with increasing pulse amplitude, but the direct CX experiences this break down at a much shorter gate time.  We note that reducing the gate duration below that which minimizes two-gate error as measured by randomized benchmarking can increase the HOP of a QV circuit, showing 
the importance of balancing circuit optimization with gate optimization.  For our successful demonstration of QV64 we used a direct CX gate duration of $\SI{199}{\nano\second}$, which is shorter than that which minimizes the two-qubit gate error.

\section{State Initialization and Readout}\label{sec:esp}
Qubit-state initialization to a fiducial simple state and qubit-specific measurement are two out of five (plus two) necessary DiVincenzo criteria for quantum computation~\cite{DiVincenzo2000}. While certain metrics are designed specifically to be insensitive to ``state preparation and measurement" (SPAM) errors, e.g. randomized benchmarking~\cite{Knill08, Magesan2011} and gate set tomography~\cite{Merkel2013,Blume-Kohout2013}, 
quantum volume was developed as a holistic system measure and hence is sensitive to SPAM-errors. \\

\begin{figure}
\includegraphics[width=8.6cm]{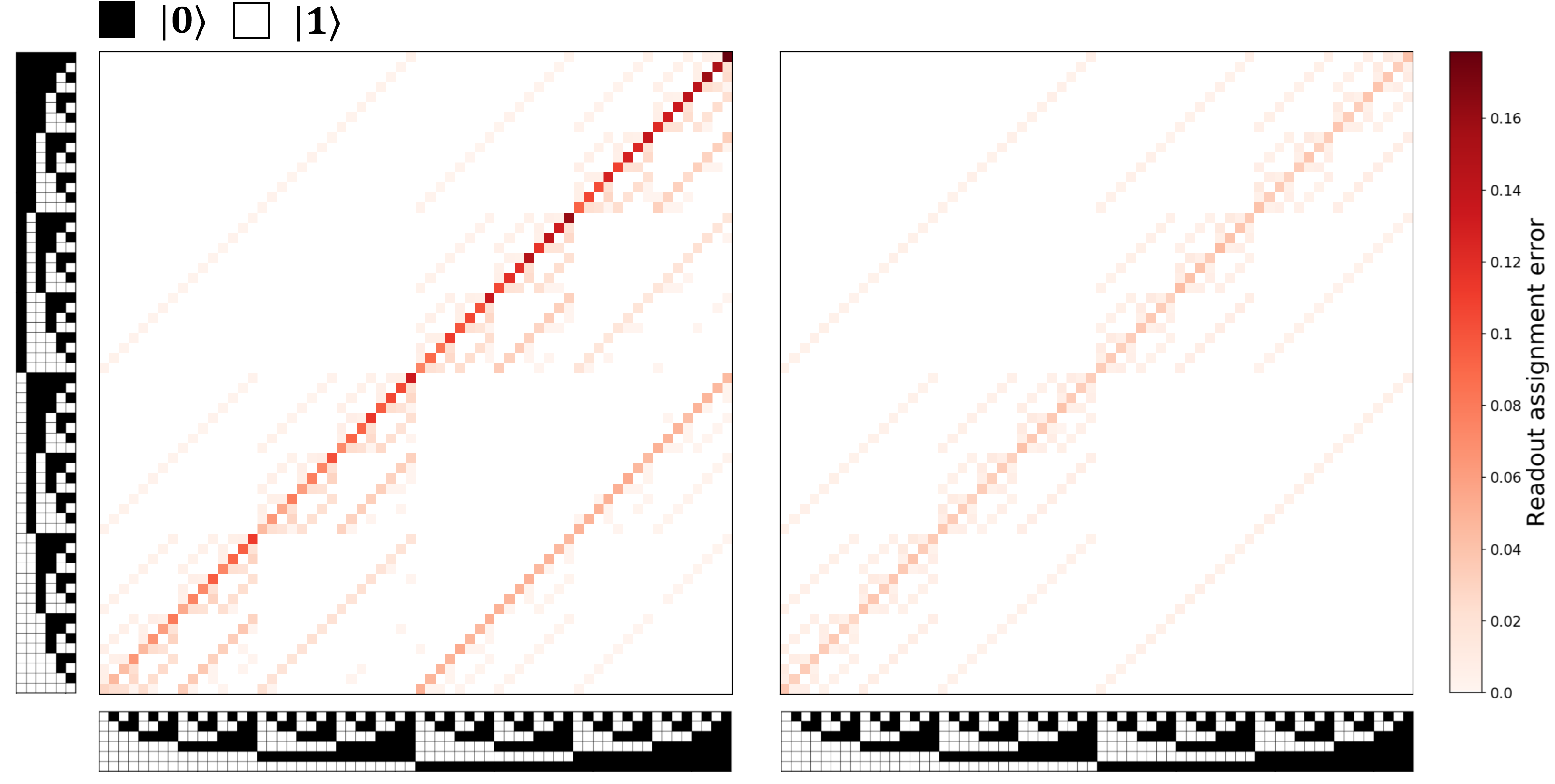}
\caption{\label{amatrix}Comparison of readout assignment-matrices. Color map indicates assignment error. Y-axis: Prepared six-qubit state vector encoded in black ($\ket{0}$) and white ($\ket{1}$). X-axis: Assigned six-qubit state vector. Left matrix: Standard procedure of the deployed system with $\ket{000000}$-state reset error  $\mathcal{E}_\mathrm{RS}=2.8\times10^{-2}$ and a total assignment error $\mathcal{E}_\mathrm{SP}=0.10$. Right matrix: State-of-the-art excited state promotion (ESP) readout with $\mathcal{E}_\mathrm{ESP}=3.5\times10^{-2}$ and $\mathcal{E}_\mathrm{RS}=3.7\times10^{-2}$}
\end{figure}

In its simplest form, qubit initialization or reset is done passively by waiting multiple $\mathrm{T}_1$ relaxation times before every new computational cycle in order to let the qubit thermalize with its surrounding bath. With ever-increasing coherence times, thermal relaxation protocols impractically limit the computational repetition rate. Various active reset schemes have been proposed and experimentally demonstrated~\cite{Magnard2018, Egger2018}. IBM Quantum systems implement a similar unconditional reset scheme~\cite{inprep_reset}. By measuring the readout matrix (Fig.~\ref{amatrix}(a)) we can infer a reset error of $\mathcal{E}_\mathrm{RS}=2.8\times10^{-2}$ for the six-qubit ground state $\ket{0\dots0}$.\\

Single qubits are dispersively read out by transversely coupled transmission line cavities~\cite{Gambetta2008}. The I-Q trajectories of each measurement signal are integrated with a filter function weighting the initial signal more heavily, hence reducing the sensitivity to $T_1$ events during measurement\cite{Ryan15}. The signal is amplified with a quantum limited travelling wave parametric amplifier followed by a classical amplification chain. This standard procedure (SP) for typically deployed systems gives a total assignment error of $\mathcal{E}_\mathrm{SP}=0.10$ for all $2^6$ states.\\

In order to further boost readout we have implemented excited state promotion (ESP) by applying an additional $\pi$-pulse between the first and second excited transmon states $\ket{1}\rightarrow\ket{f}$ before each measurement pulse~\cite{Mallet2009, Elder2020}, where $\ket{f}$ is the second excited transmon state. The advantage of this population transfer is threefold.  Firstly, the dispersive $\chi$-shift between $\ket{0}\leftrightarrow\ket{f}$ is stronger leading to a larger separation of the signals in the I-Q plane. Secondly, even though the $\ket{f}$-state has a lifetime half of the $\ket{1}$-state~\cite{Peterer2015}, the qubit excitation has to decay twice $\ket{f}\rightarrow\ket{1}\rightarrow\ket{0}$ (while a two-photon decay $\ket{f}\rightarrow\ket{0}$ is strongly suppressed~\cite{Koch2007} ). This scheme effectively extends the $\ket{1}$ qubit-state lifetime and further reduces false $\ket{0}$ assignment due to $T_1$ decays. Lastly, the relaxation through an intermediate state leads to a sub-Poissonian relaxations statistics improving the readout error even in the absence of an increase in signal-to-noise ratio \cite{D'Anjou2017}. State discrimination is set with a linear discriminant analysis (LDA) between the states $\ket{0}$ and $\ket{f}$ in the I-Q plane. In order to reset the extended qutrit system, we adapt our reset protocol in the following way: reset - $\pi_{\ket{f}\rightarrow\ket{1}}$ - reset. This state-of-the-art readout reduces the total assignment error to $\mathcal{E}_\mathrm{ESP}=3.5\times10^{-2}$ with an initialization error of $\mathcal{E}_\mathrm{RS}=3.7\times10^{-2}$, measured with the assignment matrix (Fig.~\ref{amatrix}(b)).

\section{Conclusion}\label{sec:con}

In this paper we have shown an improvement in the quantum volume of a state-of-the-art superconducting quantum system. We reached a quantum volume of 64 through a combination of four factors: improvements of the Qiskit compiler, refinements to the two-qubit gate and its calibration, addition of dynamical decoupling to mitigate noise affecting idle qubits, and introduction of excited state promoted readout. The last three techniques were developed by having lower-in-the-stack access to how the pulses and gates that comprise quantum circuits are defined before being sent to control the qubits. Furthermore, we note that optimizing the fidelity of quantum circuits is not equivalent to optimizing the gates and confirms the need for circuit benchmarks like quantum volume.  This type of hardware-aware approach to make improvements to circuit performance is a hallmark of the current era of noisy quantum systems which we expect to continue until we can achieve error rates in the range of $10^{-4}$. 

\section*{Acknowledgement}
We thank all those that contributed to the hardware system delivery and implementation, including the IBM Microelectronics Research Laboratory and Central Scientific Services teams as well as the worldwide team who designed, built and tested the custom control electronics. We further acknowledge all the work from the broader IBM Quantum team who helped this effort across the full stack. 

\bibliographystyle{unsrt}
\bibliography{bibliography.bib}

\end{document}